\documentclass[a4paper,11pt]{article}

\makeatletter
\gdef\@fpheader{   }
\gdef\@journal{jhep}

\RequirePackage{amsmath}
\RequirePackage{amssymb}
\RequirePackage{epsfig}
\RequirePackage{CJK}
\RequirePackage{graphicx}
\RequirePackage[numbers,sort&compress]{natbib}
\RequirePackage{color}
\RequirePackage[dvipdfmx
,colorlinks=true
,urlcolor=blue
,anchorcolor=blue
,citecolor=blue
,filecolor=blue
,linkcolor=blue
,menucolor=blue
,pagecolor=blue
,linktocpage=true
,pdfproducer=medialab
,pdfa=true
,CJKbookmarks
]{hyperref}

\newif\ifnotoc\notocfalse
\newif\ifemailadd\emailaddfalse
\newif\iftoccontinuous\toccontinuousfalse

\def\@subheader{\@empty}
\def\@keywords{\@empty}
\def\@abstract{\@empty}
\def\@xtum{\@empty}
\def\@dedicated{\@empty}
\def\@arxivnumber{\@empty}
\def\@collaboration{\@empty}
\def\@collaborationImg{\@empty}
\def\@proceeding{\@empty}
\def\@preprint{\@empty}

\newcommand{\subheader}[1]{\gdef\@subheader{#1}}
\newcommand{\keywords}[1]{\if!\@keywords!\gdef\@keywords{#1}\else%
\PackageWarningNoLine{\jname}{Keywords already defined.\MessageBreak Ignoring last definition.}\fi}
\renewcommand{\abstract}[1]{\gdef\@abstract{#1}}
\newcommand{\dedicated}[1]{\gdef\@dedicated{#1}}
\newcommand{\arxivnumber}[1]{\gdef\@arxivnumber{#1}}
\newcommand{\proceeding}[1]{\gdef\@proceeding{#1}}
\newcommand{\xtumfont}[1]{\textsc{#1}}
\newcommand{\correctionref}[3]{\gdef\@xtum{\xtumfont{#1} \href{#2}{#3}}}
\newcommand\jname{JHEP}
\newcommand\acknowledgments{\section*{Acknowledgments}}

\newcommand\preprint[1]{\gdef\@preprint{\hfill #1}}

\newcommand\note[2][]{%
\if!#1!%
\stepcounter{footnote}\footnotetext{#2}%
\else%
{\renewcommand\thefootnote{#1}%
\footnotetext{#2}}%
\fi}

\newtoks\auth@toks
\renewcommand{\author}[2][]{%
  \if!#1!%
    \auth@toks=\expandafter{\the\auth@toks#2\ }%
  \else
    \auth@toks=\expandafter{\the\auth@toks#2$^{#1}$\ }%
  \fi
}

\newtoks\affil@toks\newif\ifaffil\affilfalse
\newcommand{\affiliation}[2][]{%
\affiltrue
  \if!#1!%
    \affil@toks=\expandafter{\the\affil@toks{\item[]#2}}%
  \else
    \affil@toks=\expandafter{\the\affil@toks{\item[$^{#1}$]#2}}%
  \fi
}

\newtoks\email@toks\newcounter{email@counter}%
\newcommand{\emailAdd}[1]{%
\emailaddtrue%
\ifnum\theemail@counter>0\email@toks=\expandafter{\the\email@toks, \@email{#1}}%
\else\email@toks=\expandafter{\the\email@toks\@email{#1}}%
\fi\stepcounter{email@counter}}
\newcommand{\@email}[1]{\href{mailto:#1}{\tt #1}}

\newcommand*\collaboration[1]{\gdef\@collaboration{#1}}
\newcommand*\collaborationImg[2][]{\gdef\@collaborationImg{#2}}

\newcommand\afterLogoSpace{\smallskip}
\newcommand\afterSubheaderSpace{\vskip3pt plus 2pt minus 1pt}
\newcommand\afterProceedingsSpace{\vskip21pt plus0.4fil minus15pt}
\newcommand\afterTitleSpace{\vskip23pt plus0.06fil minus13pt}
\newcommand\afterRuleSpace{\vskip23pt plus0.06fil minus13pt}
\newcommand\afterCollaborationSpace{\vskip3pt plus 2pt minus 1pt}
\newcommand\afterCollaborationImgSpace{\vskip3pt plus 2pt minus 1pt}
\newcommand\afterAuthorSpace{\vskip5pt plus4pt minus4pt}
\newcommand\afterAffiliationSpace{\vskip3pt plus3pt}
\newcommand\afterEmailSpace{\vskip16pt plus9pt minus10pt\filbreak}
\newcommand\afterXtumSpace{\par\bigskip}
\newcommand\afterAbstractSpace{\vskip16pt plus9pt minus13pt}
\newcommand\afterKeywordsSpace{\vskip16pt plus9pt minus13pt}
\newcommand\afterArxivSpace{\vskip3pt plus0.01fil minus10pt}
\newcommand\afterDedicatedSpace{\vskip0pt plus0.01fil}
\newcommand\afterTocSpace{\bigskip\medskip}
\newcommand\afterTocRuleSpace{\bigskip\bigskip}
\newlength{\affiliationsSep}
\newcommand\beforetochook{\pagestyle{myplain}\pagenumbering{roman}}

\DeclareFixedFont\trfont{OT1}{phv}{b}{sc}{11}

\renewcommand\maketitle{
\pagestyle{empty}
\thispagestyle{titlepage}
\setcounter{page}{0}
\noindent{\small\scshape\@fpheader}\@preprint\par
\afterLogoSpace
\if!\@subheader!\else\noindent{\trfont{\@subheader}}\fi
\afterSubheaderSpace
\if!\@proceeding!\else\noindent{\sc\@proceeding}\fi
\afterProceedingsSpace
{\LARGE\flushleft\sffamily\bfseries\@title\par}
\afterTitleSpace
\hrule height 1.5\p@%
\afterRuleSpace
\if!\@collaboration!\else
{\Large\bfseries\sffamily\raggedright\@collaboration}\par
\afterCollaborationSpace
\fi
\if!\@collaborationImg!\else
{\normalsize\bfseries\sffamily\raggedright\@collaborationImg}\par
\afterCollaborationImgSpace
\fi
{\bfseries\raggedright\sffamily\the\auth@toks\par}
\afterAuthorSpace
\ifaffil\begin{list}{}{%
\setlength{\leftmargin}{0.28cm}%
\setlength{\labelsep}{0pt}%
\setlength{\itemsep}{\affiliationsSep}%
\setlength{\topsep}{-\parskip}}
\itshape\small%
\the\affil@toks
\end{list}\fi
\afterAffiliationSpace
\ifemailadd %% if emailadd is true
\noindent\hspace{0.28cm}\begin{minipage}[l]{.9\textwidth}
\begin{flushleft}
\textit{E-mail:} \the\email@toks
\end{flushleft}
\end{minipage}
\else %% if emailaddfalse do nothing
\PackageWarningNoLine{\jname}{E-mails are missing.\MessageBreak Plese use \protect\emailAdd\space macro to provide e-mails.}
\fi
\afterEmailSpace
\if!\@xtum!\else\noindent{\@xtum}\afterXtumSpace\fi
\if!\@abstract!\else\noindent{\renewcommand\baselinestretch{.9}\textsc{Abstract:}}\ \@abstract\afterAbstractSpace\fi
\if!\@keywords!\else\noindent{\textsc{Keywords:}} \@keywords\afterKeywordsSpace\fi
\if!\@arxivnumber!\else\noindent{\textsc{ArXiv ePrint:}} \href{http://arxiv.org/abs/\@arxivnumber}{\@arxivnumber}\afterArxivSpace\fi
\if!\@dedicated!\else\vbox{\small\it\raggedleft\@dedicated}\afterDedicatedSpace\fi
\ifnotoc\else
\iftoccontinuous\else\newpage\fi
\beforetochook\hrule
\tableofcontents
\afterTocSpace
\hrule
\afterTocRuleSpace
\fi
\setcounter{footnote}{0}
\pagestyle{myplain}\pagenumbering{arabic}
} % close the \renewcommand\maketitle{

\renewcommand{\baselinestretch}{1.1}\normalsize
\divide\@tempdima\baselineskip
\@tempcnta=\@tempdima
\skip\@mpfootins = \skip\footins
\renewcommand{\@dotsep}{10000}

\newcommand\ps@myplain{
\pagenumbering{arabic}
\renewcommand\@oddfoot{\hfill-- \thepage\ --\hfill}
\renewcommand\@oddhead{}}
\let\ps@plain=\ps@myplain

\newcommand\ps@titlepage{\renewcommand\@oddfoot{}\renewcommand\@oddhead{}}

\numberwithin{equation}{section}

\renewcommand\section{\@startsection{section}{1}{\z@}%
                                   {-3.5ex \@plus -1.3ex \@minus -.7ex}%
                                   {2.3ex \@plus.4ex \@minus .4ex}%
                                   {\normalfont\large\bfseries}}
\renewcommand\subsection{\@startsection{subsection}{2}{\z@}%
                                   {-2.3ex\@plus -1ex \@minus -.5ex}%
                                   {1.2ex \@plus .3ex \@minus .3ex}%
                                   {\normalfont\normalsize\bfseries}}
\renewcommand\subsubsection{\@startsection{subsubsection}{3}{\z@}%
                                   {-2.3ex\@plus -1ex \@minus -.5ex}%
                                   {1ex \@plus .2ex \@minus .2ex}%
                                   {\normalfont\normalsize\bfseries}}
\renewcommand\paragraph{\@startsection{paragraph}{4}{\z@}%
                                   {1.75ex \@plus1ex \@minus.2ex}%
                                   {-1em}%
                                   {\normalfont\normalsize\bfseries}}
\renewcommand\subparagraph{\@startsection{subparagraph}{5}{\parindent}%
                                   {1.75ex \@plus1ex \@minus .2ex}%
                                   {-1em}%
                                   {\normalfont\normalsize\bfseries}}

\def\fnum@figure{\textbf{\figurename\nobreakspace\thefigure}}
\def\fnum@table{\textbf{\tablename\nobreakspace\thetable}}

\long\def\@makecaption#1#2{%
  \vskip\abovecaptionskip
  \sbox\@tempboxa{\small #1. #2}%
  \ifdim \wd\@tempboxa >\hsize
    \small #1. #2\par
  \else
    \global \@minipagefalse
    \hb@xt@\hsize{\hfil\box\@tempboxa\hfil}%
  \fi
  \vskip\belowcaptionskip}

\renewenvironment{thebibliography}[1]{%
\begin{oldthebibliography}{#1}%
\small%
\raggedright%
\setlength{\itemsep}{5pt plus 0.2ex minus 0.05ex}%
}%
{%
\end{oldthebibliography}%
}

\makeatother

\usepackage[T1]{fontenc} % if needed
\usepackage{romannum} 
\usepackage{CJK}

\title{{\boldmath Scalar scattering in Schwarzschild spacetime: Integral equation method}}
\author[a,b]{Wen-Du Li,}
\author[a,b]{Yu-Zhu Chen,}
\author[b*]{and Wu-Sheng Dai}\note{daiwusheng@tju.edu.cn.}

\affiliation[a]{Theoretical Physics Division, Chern Institute of Mathematics, Nankai University, Tianjin, 300071, P. R. China}
\affiliation[b]{Department of Physics, Tianjin University, Tianjin 300350, P.R. China}

\abstract{An integral equation method for scalar scattering in Schwarzschild spacetime
is constructed. The zeroth-order and first-order scattering phase shift is obtained.
}

\keywords{Scattering, Schwarzschild spacetime, Scalar field, Integral equation method}

\begin{document} %正文开始
\begin{CJK*}{GBK}{song}
\maketitle %生成题目

\flushbottom
%(正文开始) ――――――――――――――――――――――――――――――――――――――――――――――――――

\section{Introduction}

Scattering in a curved spacetime plays an important role in gravity theory
\cite{futterman1987scattering,pike2002scattering}. There are many studies on
scattering \cite{koyama2001asymptotic,macedo2014absorption,okawa2011super}.
Approximate methods are important in the study of black-hole scattering, such
as the Born approximation \cite{batic2012born} and the WKB approximation
\cite{cho2005wkb}. Various kinds of fields, e.g., scalar fields
\cite{kuchiev2004scattering}, spinor fields
\cite{dolan2006fermion,doran2002perturbation,ahn2008black,giammatteo2005dirac,cotuaescu2016partial}%
, and vector fields
\cite{rosa2012massive,batic2012orbiting,cardoso2001quasinormal} scattered on
black holes are systematically studied. Besides the Schwarzschild spacetime,
scattering in other kinds of spacetime are also discussed, such as\ the
Reissner-Nordstr\"{o}m spacetime
\cite{benone2014absorption,crispino2009scattering,Cotaescu2016}, the Kerr
spacetime \cite{glampedakis2001scattering,brito2013massive}, and a deformed
non-rotating black hole \cite{pei2015scattering}. Some exact results are also
obtained \cite{fiziev2011application,vieira2016confluent,vieira2014exact}.

Solving a scattering problem on the Schwarzschild spacetime is to solve the
scattering solution of the scalar field equation with the Schwarzschild
metric. An effective way to solve a differential equation is to convert the
differential equation to an integral equation with the help of the Green
function. The integral equation then can be solved\ by the iterative method
\cite{courant2008methods}. The radial equation in the scalar scattering in the
Schwarzschild spacetime has both second-order derivative terms and first-order
derivative terms, and moreover, there is also a singularity on the horizon. In
this paper, we develop an integral equation method for solving the scattering
problem of a massive scalar particle in the Schwarzschild spacetime.

Using the integral equation method, we calculate the zeroth-order and
first-order contributions of the scattering wave function and the scattering
phase shift. The integral equation method constructed in\quad the present
paper is a systematic method for solving scattering in curved spacetime and in
principle can be applied to other scattering problems in gravity theory.

In section \ref{equation}, we construct the integral equation for a scalar
field in the Schwarzschild spacetime. In section \ref{SPS}, we calculate the
phase shift by solving the integral equation. The conclusions are summarized
in section \ref{conclusion}.

\section{Integral equation \label{equation}}

In this section, we convert the radial differential equation of a scalar field
in the Schwarzschild spacetime into an integral equation. Then we solve the
scattering wave function and the scattering phase shift from this integral equation.

\subsection{Integral equation}

To solve a\ scalar scattering problem in the Schwarzschild spacetime,
technically speaking, is to solve the scalar equation $\left(  \frac{1}%
{\sqrt{-g}}\frac{\partial}{\partial x^{\mu}}\sqrt{-g}g^{\mu\nu}\frac{\partial
}{\partial x^{\nu}}-\mu^{2}\right)  \Phi=0$ under the Schwarzschild metric
$ds^{2}=-\left(  1-\frac{2M}{r}\right)  dt^{2}+\left(  1-\frac{2M}{r}\right)
^{-1}dr^{2}+r^{2}d\theta^{2}+r^{2}\sin^{2}\theta d\phi^{2}$ with $\mu$ the
mass of the particle \cite{pike2002scattering}. The corresponding radial
equation is%
\begin{equation}
\left[  \frac{1}{r^{2}}\left(  1-\frac{2M}{r}\right)  \frac{d}{dr}r^{2}\left(
1-\frac{2M}{r}\right)  \frac{d}{dr}+\omega^{2}-\left(  1-\frac{2M}{r}\right)
\mu^{2}-\left(  1-\frac{2M}{r}\right)  \frac{l\left(  l+1\right)  }{r^{2}%
}\right]  \phi_{l}\left(  r\right)  =0, \label{radialeq}%
\end{equation}
where\ $\phi_{l}\left(  r\right)  $ is the radial wave function and $r\geq2M$.

Introducing $u_{l}\left(  r\right)  $ by
\begin{equation}
\phi_{l}\left(  r\right)  =\frac{u_{l}\left(  r\right)  }{r} \label{xiandu}%
\end{equation}
and substituting Eq. (\ref{xiandu}) into Eq. (\ref{radialeq}) with a variable
substitution $\rho=r/\left(  2M\right)  $ give an equation of $u_{l}\left(
\rho\right)  $:%
\begin{equation}
\left(  1-\frac{1}{\rho}\right)  \frac{d}{d\rho}\left(  1-\frac{1}{\rho
}\right)  \frac{d}{d\rho}u_{l}\left(  \rho\right)  +\left\{  \left(
2M\eta\right)  ^{2}-\left(  1-\frac{1}{\rho}\right)  \left[  \frac{l\left(
l+1\right)  }{\rho^{2}}+\frac{1}{\rho^{3}}\right]  +\frac{\left(
2M\mu\right)  ^{2}}{\rho}\right\}  u_{l}\left(  \rho\right)  =0,
\label{radialequrho}%
\end{equation}
where $\eta=\sqrt{\omega^{2}-\mu^{2}}$. By introducing an effective potential%
\begin{equation}
V_{l}^{eff}\left(  \rho\right)  =\left(  1-\frac{1}{\rho}\right)  \left[
\frac{l\left(  l+1\right)  }{\rho^{2}}+\frac{1}{\rho^{3}}\right]
-\frac{\left(  2M\mu\right)  ^{2}}{\rho}, \label{Veff}%
\end{equation}
we rewrite Eq. (\ref{radialequrho}) as%
\begin{equation}
\left(  1-\frac{1}{\rho}\right)  \frac{d}{d\rho}\left(  1-\frac{1}{\rho
}\right)  \frac{d}{d\rho}u_{l}\left(  \rho\right)  +\left(  2M\eta\right)
^{2}u_{l}\left(  \rho\right)  =V_{l}^{eff}\left(  \rho\right)  u_{l}\left(
\rho\right)  . \label{equrho}%
\end{equation}

In order to solve the equation (\ref{equrho}) by the Green function method, we
first solve Eq. (\ref{equrho}) with $V_{l}^{eff}\left(  \rho\right)  =0$,
i.e.,%
\begin{equation}
\left(  1-\frac{1}{\rho}\right)  ^{2}\frac{d^{2}}{d\rho^{2}}y_{l}\left(
\rho\right)  +\left(  1-\frac{1}{\rho}\right)  \frac{1}{\rho^{2}}\frac
{d}{d\rho}y_{l}\left(  \rho\right)  +\left(  2M\eta\right)  ^{2}y_{l}\left(
\rho\right)  =0, \label{udd}%
\end{equation}
where $y_{l}\left(  \rho\right)  $ is $u_{l}\left(  \rho\right)  $ with
$V_{l}^{eff}\left(  \rho\right)  =0$.

For $1<\rho<\infty$, Eq. (\ref{udd}) has two linearly independent solutions:%
\begin{align}
y_{l}^{\left(  1\right)  }\left(  \rho\right)   &  =\sin\left(  2M\eta\left[
\rho+\ln\left(  \rho-1\right)  \right]  \right)  ,\label{y1}\\
y_{l}^{\left(  2\right)  }\left(  \rho\right)   &  =\cos\left(  2M\eta\left[
\rho+\ln\left(  \rho-1\right)  \right]  \right)  . \label{y2}%
\end{align}
The Green function can be constructed as%
\begin{align}
G\left(  \rho,\rho^{\prime}\right)   &  =C_{1}\left(  \rho^{\prime}\right)
y_{l}^{\left(  1\right)  }\left(  \rho\right)  +C_{2}\left(  \rho^{\prime
}\right)  y_{l}^{\left(  2\right)  }\left(  \rho\right)  ,\text{ \ }\rho
>\rho^{\prime},\label{rdyrp}\\
G\left(  \rho,\rho^{\prime}\right)   &  =0,\text{
\ \ \ \ \ \ \ \ \ \ \ \ \ \ \ \ \ \ \ \ \ \ \ \ }\rho<\rho^{\prime},
\end{align}
in order to satisfy the boundary condition that the Green function must be
finite at $\rho=1$ \cite{arfken2013mathematical}.

Continuity requires that \cite{arfken2013mathematical}%
\begin{align}
\lim_{\epsilon\rightarrow0^{+}}\left.  G\left(  \rho,\rho^{\prime}\right)
\right\vert _{\rho=\rho^{\prime}+\epsilon}  &  =\lim_{\epsilon\rightarrow
0^{+}}\left.  G\left(  \rho,\rho^{\prime}\right)  \right\vert _{\rho
=\rho^{\prime}-\epsilon},\label{GeqG}\\
\lim_{\epsilon\rightarrow0^{+}}\left[  \left.  \frac{\partial}{\partial\rho
}G\left(  \rho,\rho^{\prime}\right)  \right\vert _{\rho=\rho^{\prime}%
+\epsilon}-\left.  \frac{\partial}{\partial\rho}G\left(  \rho,\rho^{\prime
}\right)  \right\vert _{\rho=\rho^{\prime}-\epsilon}\right]   &  =\frac
{1}{\left(  1-1/\rho\right)  ^{2}}. \label{dGeqdG}%
\end{align}
Then we have%
\begin{align}
&  C_{1}\left(  \rho^{\prime}\right)  \sin\left(  2M\eta\left[  \rho^{\prime
}+\ln\left(  \rho^{\prime}-1\right)  \right]  \right)  +C_{2}\left(
\rho^{\prime}\right)  \cos\left(  2M\eta\left[  \rho^{\prime}+\ln\left(
\rho^{\prime}-1\right)  \right]  \right)  =0,\label{hslx}\\
&  2M\eta\left(  1+\frac{1}{\rho^{\prime}-1}\right)  C_{1}\left(  \rho
^{\prime}\right)  \cos\left(  2M\eta\left[  \rho^{\prime}+\ln\left(
\rho^{\prime}-1\right)  \right]  \right) \nonumber\\
&  -2M\eta\left(  1+\frac{1}{\rho^{\prime}-1}\right)  C_{2}\left(
\rho^{\prime}\right)  \sin\left(  2M\eta\left[  \rho^{\prime}+\ln\left(
\rho^{\prime}-1\right)  \right]  \right)  =\frac{1}{\left(  1-1/\rho\right)
^{2}}. \label{dslx}%
\end{align}

Solving $C_{1}\left(  \rho^{\prime}\right)  $ and $C_{2}\left(  \rho^{\prime
}\right)  $ from Eqs. (\ref{hslx}) and (\ref{dslx}) and substituting
$C_{1}\left(  \rho^{\prime}\right)  $, $C_{2}\left(  \rho^{\prime}\right)  $,
and Eqs. (\ref{y1}) and (\ref{y2}) into Eq. (\ref{rdyrp}) give the Green
function,%
\begin{align}
G\left(  \rho,\rho^{\prime}\right)  =  &  \frac{\rho^{\prime}\cos\left(
2M\eta\left[  \rho^{\prime}+\ln\left(  \rho^{\prime}-1\right)  \right]
\right)  }{2M\eta\left(  \rho^{\prime}-1\right)  }\sin\left(  2M\eta\left[
\rho+\ln\left(  \rho-1\right)  \right]  \right) \nonumber\\
&  -\frac{\rho^{\prime}\sin\left(  2M\eta\left[  \rho^{\prime}+\ln\left(
\rho^{\prime}-1\right)  \right]  \right)  }{2M\eta\left(  \rho^{\prime
}-1\right)  }\cos\left(  2M\eta\left[  \rho+\ln\left(  \rho-1\right)  \right]
\right)  ,\text{ \ }\rho>\rho^{\prime}. \label{Greenfunction}%
\end{align}

In order to construct the general solution of the inhomogeneous equation
(\ref{equrho}), we start with the\ general solution of the corresponding
homogeneous equation which is the inhomogeneous equation (\ref{equrho})
without the effective potential (\ref{Veff}). The general solution of the
homogeneous equation is $Ay_{l}^{\left(  1\right)  }\left(  \rho\right)
+By_{l}^{\left(  2\right)  }\left(  \rho\right)  $. Then the general solution
of the inhomogeneous equation (\ref{equrho}) can be constructed by the general
solution of the homogeneous equation and the Green function $G\left(
\rho,\rho^{\prime}\right)  $ \cite{arfken2013mathematical}. Concretely, by the
Green function (\ref{Greenfunction}), we can establish an integral equation
for $u_{l}\left(  \rho\right)  $:%
\begin{align}
u_{l}\left(  \rho\right)   &  =Ay_{l}^{\left(  1\right)  }\left(  \rho\right)
+By_{l}^{\left(  2\right)  }\left(  \rho\right)  +\int_{1}^{\rho}G\left(
\rho,\rho^{\prime}\right)  V_{l}^{eff}\left(  \rho^{\prime}\right)
u_{l}\left(  \rho^{\prime}\right)  d\rho^{\prime}\nonumber\\
&  =A\sin\left(  2M\eta\left[  \rho+\ln\left(  \rho-1\right)  \right]
\right)  +B\cos\left(  2M\eta\left[  \rho+\ln\left(  \rho-1\right)  \right]
\right) \nonumber\\
&  +\frac{\sin\left(  2M\eta\left[  \rho+\ln\left(  \rho-1\right)  \right]
\right)  }{2M\eta}\int_{1}^{\rho}\frac{\cos\left(  2M\eta\left[  \rho^{\prime
}+\ln\left(  \rho^{\prime}-1\right)  \right]  \right)  }{\rho^{\prime}-1}%
V_{l}^{eff}\left(  \rho^{\prime}\right)  u_{l}\left(  \rho^{\prime}\right)
\rho^{\prime}d\rho^{\prime}\nonumber\\
&  -\frac{\cos\left(  2M\eta\left[  \rho+\ln\left(  \rho-1\right)  \right]
\right)  }{2M\eta}\int_{1}^{\rho}\frac{\sin\left(  2M\eta\left[  \rho^{\prime
}+\ln\left(  \rho^{\prime}-1\right)  \right]  \right)  }{\rho^{\prime}-1}%
V_{l}^{eff}\left(  \rho^{\prime}\right)  u_{l}\left(  \rho^{\prime}\right)
\rho^{\prime}d\rho^{\prime}; \label{uoutside}%
\end{align}
or, equivalently,%
\begin{align}
u_{l}\left(  \rho\right)  =  &  A\sin\left(  2M\eta\left[  \rho+\ln\left(
\rho-1\right)  \right]  \right)  +B\cos\left(  2M\eta\left[  \rho+\ln\left(
\rho-1\right)  \right]  \right) \nonumber\\
&  -\frac{1}{2M\eta}\int_{1}^{\rho}d\rho^{\prime}\frac{\rho^{\prime}}%
{\rho^{\prime}-1}\sin\left(  2M\eta\left[  \rho^{\prime}-\rho+\ln\left(
\frac{\rho^{\prime}-1}{\rho-1}\right)  \right]  \right)  V_{l}^{eff}\left(
\rho^{\prime}\right)  u_{l}\left(  \rho^{\prime}\right)  . \label{bhoutu}%
\end{align}

\subsection{Boundary condition at horizon}

In the Schwarzschild spacetime, there is a boundary condition at the horizon
$r=2M$
\cite{choudhury2004quasinormal,damour1976black,nollert1993quasinormal,liu2014scattering}%
\begin{equation}
\phi_{l}\left(  r\right)  \overset{r\rightarrow2M}{\sim}e^{\pm i\omega
r_{\ast}} \label{asyb2M}%
\end{equation}
with $r_{\ast}=r+2M\ln\left\vert \frac{r}{2M}-1\right\vert $ the tortoise
coordinate. This boundary condition can be equivalently expressed as%

\begin{equation}
u_{l}\left(  \rho\right)  \overset{\rho\rightarrow1^{+}}{\sim}e^{\pm
i2M\eta\rho_{\ast}} \label{BCuro}%
\end{equation}
with the tortoise coordinate $\rho_{\ast}=\int d\rho\frac{1}{1-1/\rho}%
=\rho+\ln\left(  \rho-1\right)  $ \cite{frolov2012black}.

The integral equation (\ref{bhoutu}) can be rewritten as%
\begin{equation}
u_{l}\left(  \rho\right)  =A\sin\left(  2M\eta\rho_{\ast}\right)
+B\cos\left(  2M\eta\rho_{\ast}\right)  -\frac{1}{2M\eta}\int_{1}^{\rho}%
\sin\left(  2M\eta\left(  \rho_{\ast}^{\prime}-\rho_{\ast}\right)  \right)
V_{l}^{eff}\left(  \rho^{\prime}\right)  u_{l}\left(  \rho^{\prime}\right)
d\rho_{\ast}^{\prime}. \label{utortoiseouter}%
\end{equation}

Near the outer horizon, i.e., $\rho\rightarrow1^{+}$ (corresponding to
$r\rightarrow2M$), Eq. (\ref{utortoiseouter}) reduces to%
\begin{equation}
\left.  u_{l}\left(  \rho\right)  \right\vert _{\rho\rightarrow1^{+}}%
=C\lim_{\rho\rightarrow1^{+}}\sin\left(  2M\eta\rho_{\ast}+\phi\right)  ,
\label{uoutside2m}%
\end{equation}
where $\cos\phi=A/\sqrt{A^{2}+B^{2}}$, $\sin\phi=B/\sqrt{A^{2}+B^{2}}$, and
$C=\sqrt{A^{2}+B^{2}}$. The superscript $+$ denotes that $\rho$ tends to the
horizon from outside.

Rewrite\ the wave function (\ref{uoutside2m}) as%
\begin{equation}
\left.  u_{l}\left(  \rho\right)  \right\vert _{\rho\rightarrow1^{+}}=\frac
{C}{2i}\lim_{\rho\rightarrow1^{+}}\left[  e^{i\left(  2M\eta\rho_{\ast}%
+\phi\right)  }-e^{-i\left(  2M\eta\rho_{\ast}+\phi\right)  }\right]  ,
\label{uoutsideexp}%
\end{equation}
which includes two parts:%
\begin{align}
\left.  u_{l}\left(  \rho\right)  \right\vert _{\rho\rightarrow1^{+}}^{out}
&  =\frac{C}{2i}e^{i\left(  2M\eta\rho_{\ast}+\phi\right)  },\text{
\ \ outgoing wave,}\label{uoutsideexpmu}\\
\left.  u_{l}\left(  \rho\right)  \right\vert _{\rho\rightarrow1^{+}}^{in}  &
=\frac{C}{2i}e^{-i\left(  2M\eta\rho_{\ast}+\phi\right)  },\text{ \ ingoing
wave.}%
\end{align}
Clearly, this satisfies the boundary condition (\ref{BCuro}).

\section{Scattering phase shift \label{SPS}}

In this section, we calculate the scattering phase shift for high energy
scattering, i.e., $\mu/\eta\ll1$, based on the integral equation constructed above.

\subsection{Scattering phase shift}

The radial equation, Eq. (\ref{radialequrho}), under the replacement
$2\rho-1\rightarrow\rho$, is the Heun equation. The $\rho\rightarrow\infty$
asymptotic solution, for $\mu/\eta\ll1$, is \cite{ronveaux1995heun}%
\begin{align}
&  u_{l}\left(  \rho\right)  \overset{\rho\rightarrow\infty}{\sim}\sin\left(
2M\eta\left(  \rho+\ln\left(  \rho-1\right)  \right)  +\delta_{l}-\frac{l\pi
}{2}-\eta M+2M\eta\ln2\right)  \nonumber\\
&  =\sin\left(  2M\eta\left[  \rho+\ln\left(  \rho-1\right)  \right]  \right)
\cos\left(  \delta_{l}+\Delta\left(  \eta,M\right)  \right)  +\cos\left(
2M\eta\left[  \rho+\ln\left(  \rho-1\right)  \right]  \right)  \sin\left(
\delta_{l}+\Delta\left(  \eta,M\right)  \right)  ,\label{uinf1}%
\end{align}
where $\rho=\frac{r}{2M}$ and $\Delta\left(  \eta,M\right)  =-\frac{l\pi}%
{2}-\eta M+2M\eta\ln2$ are used.

It can be seen that the asymptotic wave function, Eq. (\ref{uinf1}), is
determined by the phase shift $\delta_{l}$: once the phase shift is obtained,
the asymptotic wave function is obtained. This is the same as the
quantum-mechanical scattering: all information of the scattering wave function
is embodied in the phase shift. Therefore, in a scattering problem, the main
task is to find\ the phase shift.

In order to compare with (\ref{uinf1}), we rewrite Eq. (\ref{uoutside}) as%
\begin{align}
u_{l}\left(  \rho\right)   &  =\sin\left(  2M\eta\left[  \rho+\ln\left(
\rho-1\right)  \right]  \right)  \left[  A+\frac{1}{2M\eta}\int_{1}^{\rho
}\frac{\cos\left(  2M\eta\left[  \rho^{\prime}+\ln\left(  \rho^{\prime
}-1\right)  \right]  \right)  }{\rho^{\prime}-1}V_{l}^{eff}\left(
\rho^{\prime}\right)  u_{l}\left(  \rho^{\prime}\right)  \rho^{\prime}%
d\rho^{\prime}\right] \nonumber\\
&  +\cos\left(  2M\eta\left[  \rho+\ln\left(  \rho-1\right)  \right]  \right)
\left[  B-\frac{1}{2M\eta}\int_{1}^{\rho}\frac{\sin\left(  2M\eta\left[
\rho^{\prime}+\ln\left(  \rho^{\prime}-1\right)  \right]  \right)  }%
{\rho^{\prime}-1}V_{l}^{eff}\left(  \rho^{\prime}\right)  u_{l}\left(
\rho^{\prime}\right)  \rho^{\prime}d\rho^{\prime}\right]  .
\end{align}
and, then, take $\rho\rightarrow\infty$ asymptotics:%
\begin{align}
u_{l}\left(  \rho\right)   &  \overset{\rho\rightarrow\infty}{\sim}%
\alpha\left(  \eta,M\right)  \sin\left(  2M\eta\left[  \rho+\ln\left(
\rho-1\right)  \right]  \right)  +\beta\left(  \eta,M\right)  \cos\left(
2M\eta\left[  \rho+\ln\left(  \rho-1\right)  \right]  \right) \nonumber\\
&  =C\sin\left(  2M\eta\left[  \rho+\ln\left(  \rho-1\right)  \right]
+\phi\right)  , \label{uinf2}%
\end{align}
where $\alpha\left(  \eta,M\right)  =A+\frac{1}{2M\eta}\int_{1}^{\infty}%
\frac{\cos\left(  2M\eta\left[  \rho^{\prime}+\ln\left(  \rho^{\prime
}-1\right)  \right]  \right)  }{\rho^{\prime}-1}V_{l}^{eff}\left(
\rho^{\prime}\right)  u_{l}\left(  \rho^{\prime}\right)  \rho^{\prime}%
d\rho^{\prime}$, $\beta\left(  \eta,M\right)  =B-\frac{1}{2M\eta}\int%
_{1}^{\infty}\frac{\sin\left(  2M\eta\left[  \rho^{\prime}+\ln\left(
\rho^{\prime}-1\right)  \right]  \right)  }{\rho^{\prime}-1}V_{l}^{eff}\left(
\rho^{\prime}\right)  u_{l}\left(  \rho^{\prime}\right)  \rho^{\prime}%
d\rho^{\prime}$, $\cos\phi=\alpha\left(  \eta,M\right)  /\sqrt{\alpha
^{2}\left(  \eta,M\right)  +\beta^{2}\left(  \eta,M\right)  }$, $\sin
\phi=\beta\left(  \eta,M\right)  /\sqrt{\alpha^{2}\left(  \eta,M\right)
+\beta^{2}\left(  \eta,M\right)  }$, and the normalization constant
$C=\sqrt{\alpha^{2}\left(  \eta,M\right)  +\beta^{2}\left(  \eta,M\right)  }$.

Comparing Eqs. (\ref{uinf1}) and (\ref{uinf2}), we have%
\begin{align}
\tan\left(  \delta_{l}+\Delta\left(  \eta,M\right)  \right)   &  =\frac
{\beta\left(  \eta,M\right)  }{\alpha\left(  \eta,M\right)  }\nonumber\\
&  =\frac{B-\frac{1}{2M\eta}\int_{1}^{\infty}\frac{\sin\left(  2M\eta\left[
\rho^{\prime}+\ln\left(  \rho^{\prime}-1\right)  \right]  \right)  }%
{\rho^{\prime}-1}V_{l}^{eff}\left(  \rho^{\prime}\right)  u_{l}\left(
\rho^{\prime}\right)  \rho^{\prime}d\rho^{\prime}}{A+\frac{1}{2M\eta}\int%
_{1}^{\infty}\frac{\cos\left(  2M\eta\left[  \rho^{\prime}+\ln\left(
\rho^{\prime}-1\right)  \right]  \right)  }{\rho^{\prime}-1}V_{l}^{eff}\left(
\rho^{\prime}\right)  u_{l}\left(  \rho^{\prime}\right)  \rho^{\prime}%
d\rho^{\prime}}, \label{tandelta}%
\end{align}
where%
\begin{align}
\phi &  =\arctan\frac{\beta\left(  \eta,M\right)  }{\alpha\left(
\eta,M\right)  }\nonumber\\
&  =\frac{B-\frac{1}{2M\eta}\int_{1}^{\infty}\frac{\sin\left(  2M\eta\left[
\rho^{\prime}+\ln\left(  \rho^{\prime}-1\right)  \right]  \right)  }%
{\rho^{\prime}-1}V_{l}^{eff}\left(  \rho^{\prime}\right)  u_{l}\left(
\rho^{\prime}\right)  \rho^{\prime}d\rho^{\prime}}{A+\frac{1}{2M\eta}\int%
_{1}^{\infty}\frac{\cos\left(  2M\eta\left[  \rho^{\prime}+\ln\left(
\rho^{\prime}-1\right)  \right]  \right)  }{\rho^{\prime}-1}V_{l}^{eff}\left(
\rho^{\prime}\right)  u_{l}\left(  \rho^{\prime}\right)  \rho^{\prime}%
d\rho^{\prime}}%
\end{align}
is used.

Then we arrive at an expression of the scattering phase shift,%
\begin{equation}
\delta_{l}=\arctan\left(  \frac{B-\frac{1}{2M\eta}\int_{1}^{\infty}\frac
{\sin\left(  2M\eta\left[  \rho^{\prime}+\ln\left(  \rho^{\prime}-1\right)
\right]  \right)  }{\rho^{\prime}-1}V_{l}^{eff}\left(  \rho^{\prime}\right)
u_{l}\left(  \rho^{\prime}\right)  \rho^{\prime}d\rho^{\prime}}{A+\frac
{1}{2M\eta}\int_{1}^{\infty}\frac{\cos\left(  2M\eta\left[  \rho^{\prime}%
+\ln\left(  \rho^{\prime}-1\right)  \right]  \right)  }{\rho^{\prime}-1}%
V_{l}^{eff}\left(  \rho^{\prime}\right)  u_{l}\left(  \rho^{\prime}\right)
\rho^{\prime}d\rho^{\prime}}\right)  -\Delta\left(  \eta,M\right)  .
\label{arctanps}%
\end{equation}
This is a relation between the scattering phase shift $\delta_{l}$ and the
scattering wave function $u_{l}\left(  \rho\right)  $.

Now we determine the constants $A$ and $B$.

When $V_{l}^{eff}\left(  \rho\right)  =0$, Eq. (\ref{utortoiseouter}) gives%
\begin{align}
u_{l}\left(  \rho\right)   &  =A\sin\left(  2M\eta\rho_{\ast}\right)
+B\cos\left(  2M\eta\rho_{\ast}\right) \nonumber\\
&  =A\sin\left(  \eta\left[  r+2M\ln\left(  \frac{r}{2M}-1\right)  \right]
\right)  +B\cos\left(  \eta\left[  r+2M\ln\left(  \frac{r}{2M}-1\right)
\right]  \right)  .
\end{align}
For $M=0$,%
\[
u_{l}\left(  r\right)  =A\sin\left(  \eta r\right)  +B\cos\left(  \eta
r\right)
\]
and the horizon is at $r=2M=0$. The boundary condition requires that $\phi
_{l}\left(  r\right)  =\frac{u_{l}\left(  r\right)  }{r}$ must be finite at
the horizon, i.e., $u_{l}\left(  0\right)  =0$. This gives $B=0$. From Eqs.
(\ref{arctanps}) and (\ref{utortoiseouter}), we can see that the constant $A$
will be eliminated finally, i.e.,%
\begin{equation}
\delta_{l}=-\arctan\left(  \frac{\frac{1}{2M\eta}\int_{1}^{\infty}\frac
{\sin\left(  2M\eta\left[  \rho^{\prime}+\ln\left(  \rho^{\prime}-1\right)
\right]  \right)  }{\rho^{\prime}-1}V_{l}^{eff}\left(  \rho^{\prime}\right)
u_{l}\left(  \rho^{\prime}\right)  \rho^{\prime}d\rho^{\prime}}{1+\frac
{1}{2M\eta}\int_{1}^{\infty}\frac{\cos\left(  2M\eta\left[  \rho^{\prime}%
+\ln\left(  \rho^{\prime}-1\right)  \right]  \right)  }{\rho^{\prime}-1}%
V_{l}^{eff}\left(  \rho^{\prime}\right)  u_{l}\left(  \rho^{\prime}\right)
\rho^{\prime}d\rho^{\prime}}\right)  -\Delta\left(  \eta,M\right)
\label{phaseshift}%
\end{equation}
That is, the constant $A$ can take any nonzero value; here and after, we take
$A=1$:%
\begin{equation}
u_{l}\left(  \rho\right)  =\sin\left(  2M\eta\rho_{\ast}\right)  -\frac
{1}{2M\eta}\int_{1}^{\rho}\sin\left(  2M\eta\left(  \rho_{\ast}^{\prime}%
-\rho_{\ast}\right)  \right)  V_{l}^{eff}\left(  \rho^{\prime}\right)
u_{l}\left(  \rho^{\prime}\right)  d\rho_{\ast}^{\prime}. \label{wavefunction}%
\end{equation}

Next, in order to obtain the scattering phase shift by Eq. (\ref{arctanps}),
we need to iteratively solve the wave function $u_{l}\left(  \rho\right)  $.

\subsection{Zeroth-order and first-order phase shifts}

By iteratively solving the integral equation of the wave function
$u_{l}\left(  \rho\right)  $, Eq. (\ref{bhoutu}), we can obtain various orders
of $u_{l}\left(  \rho\right)  $. In this section, we solve the zeroth-order
and first-order scattering phase shifts.

\textit{Zeroth-order phase shift.} The zeroth-order contribution of Eq.
(\ref{phaseshift}) is%
\begin{equation}
\left[  \tan\left(  \delta_{l}+\Delta\left(  \eta,M\right)  \right)  \right]
^{\left(  0\right)  }=0.
\end{equation}
The zeroth-order scattering phase shift then reads%
\begin{equation}
\delta_{l}^{\left(  0\right)  }=-\Delta\left(  \eta,M\right)  =\frac{l\pi}%
{2}+\eta M-2M\eta\ln2.
\end{equation}

\textit{First-order phase shift.} The first-order phase shift $\delta
_{l}^{\left(  1\right)  }$ can be obtained by substituting the zeroth-order
scattering wave function $u_{l}^{\left(  0\right)  }\left(  \rho\right)
=\sin\left(  2M\eta\left[  \rho+\ln\left(  \rho-1\right)  \right]  \right)  $
into Eq. (\ref{phaseshift}):%
\begin{equation}
\delta_{l}^{\left(  1\right)  }=-\arctan\left(  \frac{\int_{1}^{\infty}%
\sin^{2}\left(  2M\eta\left[  \rho+\ln\left(  \rho-1\right)  \right]  \right)
\frac{1}{\rho-1}V_{l}^{eff}\left(  \rho\right)  \rho d\rho}{2\text{$M$}%
\eta+\frac{1}{2}\int_{1}^{\infty}\sin\left(  4M\eta\left[  \rho+\ln\left(
\rho-1\right)  \right]  \right)  \frac{1}{\rho-1}V_{l}^{eff}\left(
\rho\right)  \rho d\rho}\right)  . \label{delta1}%
\end{equation}
The integral in Eq. (\ref{delta1}) can be worked out analytically, but it is
too complicated to be listed here.

\section{Conclusion \label{conclusion}}

An integral equation method for solving scattering of a scalar field in the
Schwarzschild spacetime is constructed. By solving the integral equation, we
obtain the zeroth-order and first-order scattering phase shifts.

Scattering in curved spacetime is an important issue and has been discussed in
many literatures. In this letter, we calculate the zeroth-order and
first-order scattering wave functions and scattering phase shifts of scalar
scattering in the Schwarzschild spacetime. There are some authors also
consider the scalar scattering in the Schwarzschild spacetime. In Ref.
\cite{koyama2001asymptotic}, the author consider the late-time evolution in
the Schwarzschild background. In Ref. \cite{batic2012born}, the author
calculate the scattering amplitude; they also calculate the phase shift, but
the phase shift is a large $r$ Coulomb-like phase shift. In Ref.
\cite{batic2012orbiting}, the author\ concentrates on the reflection
coefficient through the asymptotic solution. Based on the confluent Heun
function, some authors calculate the quasinormal modes of nonrotating black
holes \cite{fiziev2011application}, resonant frequencies, Hawking radiation,
and scattering of scalar waves \cite{vieira2016confluent}, and the angular and
radial solutions \cite{vieira2014exact}. Other types of fields scattered in a
curved spacetime are also considered, such as spinor fields
\cite{cho2005wkb,dolan2006fermion,doran2002perturbation,ahn2008black,cotuaescu2016partial}
and vector fields \cite{rosa2012massive,cardoso2001quasinormal}. Beyond the
Schwarzschild spacetime, there are many discussions devote to other types of
spacetime, such as the Reissner--Nordstr\"{o}m spacetime
\cite{macedo2014absorption,kuchiev2004scattering,benone2014absorption,crispino2009scattering,Cotaescu2016}%
, the Kerr spacetime \cite{glampedakis2001scattering}, deformed black hole
\cite{pei2015scattering}, and the AdS spacetime \cite{giammatteo2005dirac}.

The integral equation method suggested in the present paper is a method for
solving the radial equations with various effective potentials. The field
equations in different spacetimes correspond to different effective
potentials. Besides the Schwarzschild spacetime, the integral equation method
can be applied to more general cases, such as the charged RN black hole and
the spinning Kerr black hole. Concretely, for the charged RN black hole, the
field equation can be separated into two parts: the radial equation and the
angular equation. The angular equation can be solved exactly and the solution,
the same as the Schwarzschild case, is the spherical harmonics function. Then
the remaining task is to solve the radial equation. For the spinning Kerr
black hole, the angular equation can also be separated and solved exactly, and
the solution is the confluent Heun function. Again, the remaining task then is
to solve the radial equation.

It is worthy to note that scattering by Schwarzschild spacetime is a
long-range potential scattering \cite{futterman1987scattering}, i.e., this is
an integral equation method for long-range potential scattering. The
long-range scattering is more difficult than short-range scattering
\cite{li2016exact,hod2013scattering,liu2014scattering,li2016scattering}.

The method developed in the present paper is an integral equation method for
solving scattering phase shifts. This method can be applied to the scattering
spectral method \cite{graham2009spectral,pang2012relation} and heat kernel
method
\cite{barvinsky1987beyond,barvinsky1990covariant,barvinsky1990covariant3,mukhanov2007introduction,li2015heat}%
. Furthermore, through the scattering spectral method and the heat kernel
method, the method for scattering phase shift can be used to quantum field
theory
\cite{dai2009number,dai2010approach,graham2009spectral,vassilevich2003heat}.
%\appendix
%\section{Some title}
%Please always give a title also for appendices.

\acknowledgments

We are very indebted to Dr G. Zeitrauman for his encouragement. This work is supported in part by NSF of China under Grant
No. 11575125 and No. 11675119.

%(正文结束)――――――――――――――――――――――――――――――――――――――――――――――――――

%%%%%%%%%%%%%%%%%%%参考文献%%%%%%%%%%%%%%%%%%%%%%%%%%%

%%%%%%%%%%%%%%%%%%%bibtex形式的参考文献%%%%%%%%%%%%%%%
%\bibliographystyle{JHEP} %参考文献的风格(.bst)
%\bibliography{refs} %参考文献文件(.bib)

\begin{thebibliography}{10}

%bbl放在这

\bibitem{futterman1987scattering}
J.~Futterman, F.~Handler, and R.~Matzner, {\em Scattering from Black Holes}.
\newblock Cambridge Monographs on Mathematical Physics. Cambridge University
  Press, 2009.

\bibitem{pike2002scattering}
E.~Pike and P.~Sabatier, {\em Scattering: Scattering and Inverse Scattering in
  Pure and Applied Science}.
\newblock Scattering: Scattering and Inverse Scattering in Pure and Applied
  Science. Academic Press, 2002.

\bibitem{koyama2001asymptotic}
H.~Koyama and A.~Tomimatsu, {\it Asymptotic tails of massive scalar fields in a
  schwarzschild background},  {\em Physical Review D} {\bf 64} (2001), no.~4
  044014.

\bibitem{macedo2014absorption}
C.~F. Macedo and L.~C. Crispino, {\it Absorption of planar massless scalar
  waves by bardeen regular black holes},  {\em Physical Review D} {\bf 90}
  (2014), no.~6 064001.

\bibitem{okawa2011super}
H.~Okawa, K.-i. Nakao, and M.~Shibata, {\it Is super-planckian physics visible?
  scattering of black holes in 5 dimensions},  {\em Physical Review D} {\bf 83}
  (2011), no.~12 121501.

\bibitem{batic2012born}
D.~Batic, N.~Kelkar, and M.~Nowakowski, {\it On born approximation in black
  hole scattering},  {\em The European Physical Journal C} {\bf 71} (2012),
  no.~12 1--8.

\bibitem{cho2005wkb}
H.~Cho and Y.~Lin, {\it Wkb analysis of the scattering of massive dirac fields
  in schwarzschild black-hole spacetimes},  {\em Classical and Quantum Gravity}
  {\bf 22} (2005), no.~5 775.

\bibitem{kuchiev2004scattering}
M.~Y. Kuchiev and V.~Flambaum, {\it Scattering of scalar particles by a black
  hole},  {\em Physical Review D} {\bf 70} (2004), no.~4 044022.

\bibitem{dolan2006fermion}
S.~Dolan, C.~Doran, and A.~Lasenby, {\it Fermion scattering by a schwarzschild
  black hole},  {\em Physical Review D} {\bf 74} (2006), no.~6 064005.

\bibitem{doran2002perturbation}
C.~Doran and A.~Lasenby, {\it Perturbation theory calculation of the black hole
  elastic scattering cross section},  {\em Physical Review D} {\bf 66} (2002),
  no.~2 024006.

\bibitem{ahn2008black}
D.~Ahn, Y.~Moon, R.~Mann, and I.~Fuentes-Schuller, {\it The black hole final
  state for the dirac fields in schwarzschild spacetime},  {\em Journal of High
  Energy Physics} {\bf 2008} (2008), no.~06 062.

\bibitem{giammatteo2005dirac}
M.~Giammatteo and J.~Jing, {\it Dirac quasinormal frequencies in
  schwarzschild-ads space-time},  {\em Physical Review D} {\bf 71} (2005),
  no.~2 024007.

\bibitem{cotuaescu2016partial}
I.~I. Cot{\u{a}}escu, C.~Crucean, and C.~A. Sporea, {\it Partial wave analysis
  of the dirac fermions scattered from schwarzschild black holes},  {\em The
  European Physical Journal C} {\bf 76} (2016), no.~3 102.

\bibitem{rosa2012massive}
J.~G. Rosa and S.~R. Dolan, {\it Massive vector fields on the schwarzschild
  spacetime: quasinormal modes and bound states},  {\em Physical Review D} {\bf
  85} (2012), no.~4 044043.

\bibitem{batic2012orbiting}
D.~Batic, N.~Kelkar, and M.~Nowakowski, {\it Orbiting phenomena in black hole
  scattering},  {\em Physical Review D} {\bf 86} (2012), no.~10 104060.

\bibitem{cardoso2001quasinormal}
V.~Cardoso and J.~P. Lemos, {\it Quasinormal modes of schwarzschild--anti-de
  sitter black holes: Electromagnetic and gravitational perturbations},  {\em
  Physical Review D} {\bf 64} (2001), no.~8 084017.

\bibitem{benone2014absorption}
C.~L. Benone, E.~S. de~Oliveira, S.~R. Dolan, and L.~C. Crispino, {\it
  Absorption of a massive scalar field by a charged black hole},  {\em Physical
  Review D} {\bf 89} (2014), no.~10 104053.

\bibitem{crispino2009scattering}
L.~C. Crispino, S.~R. Dolan, and E.~S. Oliveira, {\it Scattering of massless
  scalar waves by reissner-nordstr{\"o}m black holes},  {\em Physical Review D}
  {\bf 79} (2009), no.~6 064022.

\bibitem{Cotaescu2016}
I.~I. Cotaescu, C.~Crucean, and C.~A. Sporea, {\it Partial wave analysis of the
  dirac fermions scattered from reissner--nordstr{\"o}m charged black holes},
  {\em The European Physical Journal C} {\bf 76} (Jul, 2016) 413.

\bibitem{glampedakis2001scattering}
K.~Glampedakis and N.~Andersson, {\it Scattering of scalar waves by rotating
  black holes},  {\em Classical and Quantum Gravity} {\bf 18} (2001), no.~10
  1939.

\bibitem{brito2013massive}
R.~Brito, V.~Cardoso, and P.~Pani, {\it Massive spin-2 fields on black hole
  spacetimes: Instability of the schwarzschild and kerr solutions and bounds on
  the graviton mass},  {\em Physical Review D} {\bf 88} (2013), no.~2 023514.

\bibitem{pei2015scattering}
G.~Pei and C.~Bambi, {\it Scattering of particles by deformed non-rotating
  black holes},  {\em The European Physical Journal C} {\bf 75} (2015), no.~11
  1--7.

\bibitem{fiziev2011application}
P.~Fiziev and D.~Staicova, {\it Application of the confluent heun functions for
  finding the quasinormal modes of nonrotating black holes},  {\em Physical
  Review D} {\bf 84} (2011), no.~12 127502.

\bibitem{vieira2016confluent}
H.~Vieira and V.~Bezerra, {\it Confluent heun functions and the physics of
  black holes: Resonant frequencies, hawking radiation and scattering of scalar
  waves},  {\em Annals of Physics} {\bf 373} (2016) 28--42.

\bibitem{vieira2014exact}
H.~Vieira, V.~Bezerra, and C.~Muniz, {\it Exact solutions of the klein--gordon
  equation in the kerr--newman background and hawking radiation},  {\em Annals
  of Physics} {\bf 350} (2014) 14--28.

\bibitem{courant2008methods}
R.~Courant and D.~Hilbert, {\em Methods of Mathematical Physics}.
\newblock No.~v. 1. Wiley, 2008.

\bibitem{arfken2013mathematical}
G.~Arfken, H.~Weber, and F.~Harris, {\em Mathematical Methods for Physicists: A
  Comprehensive Guide}.
\newblock Elsevier Science, 2013.

\bibitem{choudhury2004quasinormal}
T.~R. Choudhury and T.~Padmanabhan, {\it Quasinormal modes in schwarzschild-de
  sitter spacetime: a simple derivation of the level spacing of the
  frequencies},  {\em Physical Review D} {\bf 69} (2004), no.~6 064033.

\bibitem{damour1976black}
T.~Damour and R.~Ruffini, {\it Black-hole evaporation in the
  klein-sauter-heisenberg-euler formalism},  {\em Physical Review D} {\bf 14}
  (1976), no.~2 332.

\bibitem{nollert1993quasinormal}
H.-P. Nollert, {\it Quasinormal modes of schwarzschild black holes: The
  determination of quasinormal frequencies with very large imaginary parts},
  {\em Physical Review D} {\bf 47} (1993), no.~12 5253.

\bibitem{liu2014scattering}
T.~Liu, W.-D. Li, and W.-S. Dai, {\it Scattering theory without large-distance
  asymptotics},  {\em Journal of High Energy Physics} {\bf 2014} (2014), no.~6
  1--12.

\bibitem{frolov2012black}
V.~Frolov and I.~Novikov, {\em Black hole physics: basic concepts and new
  developments}, vol.~96.
\newblock Springer Science \& Business Media, 2012.

\bibitem{ronveaux1995heun}
A.~Ronveaux and F.~M. Arscott, {\em Heun's differential equations}.
\newblock Oxford University Press, 1995.

\bibitem{li2016exact}
W.-D. Li and W.-S. Dai, {\it Exact solution of inverse-square-root potential v
  (r)=- $\alpha$r},  {\em Annals of Physics} {\bf 373} (2016) 207--215.

\bibitem{hod2013scattering}
S.~Hod, {\it Scattering by a long-range potential},  {\em Journal of High
  Energy Physics} {\bf 2013} (2013), no.~9 1--11.

\bibitem{li2016scattering}
W.-D. Li and W.-S. Dai, {\it Scattering theory without large-distance
  asymptotics in arbitrary dimensions},  {\em Journal of Physics A:
  Mathematical and Theoretical} {\bf 49} (2016), no.~46 465202.

\bibitem{graham2009spectral}
N.~Graham, M.~Quandt, and H.~Weigel, {\em Spectral methods in quantum field
  theory}, vol.~777.
\newblock Springer, 2009.

\bibitem{pang2012relation}
H.~Pang, W.-S. Dai, and M.~Xie, {\it Relation between heat kernel method and
  scattering spectral method},  {\em The European Physical Journal C} {\bf 72}
  (2012), no.~5 1--13.

\bibitem{barvinsky1987beyond}
A.~Barvinsky and G.~Vilkovisky, {\it Beyond the schwinger-dewitt technique:
  Converting loops into trees and in-in currents},  {\em Nuclear Physics B}
  {\bf 282} (1987) 163--188.

\bibitem{barvinsky1990covariant}
A.~Barvinsky and G.~Vilkovisky, {\it Covariant perturbation theory (ii). second
  order in the curvature. general algorithms},  {\em Nuclear Physics B} {\bf
  333} (1990), no.~2 471--511.

\bibitem{barvinsky1990covariant3}
A.~Barvinsky and G.~Vilkovisky, {\it Covariant perturbation theory (iii).
  spectral representations of the third-order form factors},  {\em Nuclear
  Physics B} {\bf 333} (1990), no.~2 512--524.

\bibitem{mukhanov2007introduction}
V.~Mukhanov and S.~Winitzki, {\em Introduction to quantum effects in gravity}.
\newblock Cambridge University Press, 2007.

\bibitem{li2015heat}
W.-D. Li and W.-S. Dai, {\it Heat-kernel approach for scattering},  {\em The
  European Physical Journal C} {\bf 75} (2015), no.~6 294.

\bibitem{dai2009number}
W.-S. Dai and M.~Xie, {\it The number of eigenstates: counting function and
  heat kernel},  {\em Journal of High Energy Physics} {\bf 2009} (2009), no.~02
  033.

\bibitem{dai2010approach}
W.-S. Dai and M.~Xie, {\it An approach for the calculation of one-loop
  effective actions, vacuum energies, and spectral counting functions},  {\em
  Journal of High Energy Physics} {\bf 2010} (2010), no.~6 1--29.

\bibitem{vassilevich2003heat}
D.~V. Vassilevich, {\it Heat kernel expansion: user's manual},  {\em Physics
  reports} {\bf 388} (2003), no.~5 279--360.

\end{thebibliography}
%\nocite{*} %若去掉注释，没有被引用的文献也被列出

%%%%%%%%%%%%%%%%%%%bbl形式的参考文献%%%%%%%%%%%%%%%%%%

%%%%%%%%%%%%%%%%%%%%%%%%%%%%%%%%%%%%%%%%%%%%%%%%%%%%%%

\end{CJK*}
\end{document}